\documentclass[reprint, superscriptaddress, twocolumn, amsmath, amssymb, aps, prb]{revtex4-2}
\usepackage{graphicx}
\usepackage{dcolumn}
\usepackage{bm}
\usepackage{placeins}
\usepackage{appendix}
\usepackage{upgreek}

\usepackage{verbatim}
\usepackage[usenames,dvipsnames]{xcolor}
 \usepackage{amsmath}
 \usepackage{amsfonts}
 \usepackage{amssymb}
 \usepackage[colorlinks=true,citecolor=blue,linkcolor=red]{hyperref}

 \usepackage{bbold}     
 \usepackage[makeroom]{cancel}  
 \usepackage{multirow}    
 \usepackage[normalem]{ulem}        
 \usepackage{array}
 \usepackage{pdfpages}
\makeatletter
 \AtBeginDocument{\let\LS@rot\@undefined}
 \makeatother

\begin{document}


\title{Conductance Quantization in PbTe Nanowires}

\author{Wenyu Song}
\email{equal contribution}
\affiliation{State Key Laboratory of Low Dimensional Quantum Physics, Department of Physics, Tsinghua University, Beijing 100084, China}

\author{Yuhao Wang}
 \email{equal contribution}
\affiliation{State Key Laboratory of Low Dimensional Quantum Physics, Department of Physics, Tsinghua University, Beijing 100084, China}

\author{Wentao Miao}
\affiliation{State Key Laboratory of Low Dimensional Quantum Physics, Department of Physics, Tsinghua University, Beijing 100084, China}

\author{Zehao Yu}
\affiliation{State Key Laboratory of Low Dimensional Quantum Physics, Department of Physics, Tsinghua University, Beijing 100084, China}

\author{Yichun Gao}
\affiliation{State Key Laboratory of Low Dimensional Quantum Physics, Department of Physics, Tsinghua University, Beijing 100084, China}

\author{Ruidong Li}
\affiliation{State Key Laboratory of Low Dimensional Quantum Physics, Department of Physics, Tsinghua University, Beijing 100084, China}

\author{Shuai Yang}
\affiliation{State Key Laboratory of Low Dimensional Quantum Physics, Department of Physics, Tsinghua University, Beijing 100084, China}

\author{Fangting Chen}
\affiliation{State Key Laboratory of Low Dimensional Quantum Physics, Department of Physics, Tsinghua University, Beijing 100084, China}

\author{Zuhan Geng}
\affiliation{State Key Laboratory of Low Dimensional Quantum Physics, Department of Physics, Tsinghua University, Beijing 100084, China}

\author{Zitong Zhang}
\affiliation{State Key Laboratory of Low Dimensional Quantum Physics, Department of Physics, Tsinghua University, Beijing 100084, China}

\author{Shan Zhang}
\affiliation{State Key Laboratory of Low Dimensional Quantum Physics, Department of Physics, Tsinghua University, Beijing 100084, China}

\author{Yunyi Zang}
\affiliation{Beijing Academy of Quantum Information Sciences, Beijing 100193, China}

\author{Zhan Cao}
\affiliation{Beijing Academy of Quantum Information Sciences, Beijing 100193, China}

\author{Dong E. Liu}
\affiliation{State Key Laboratory of Low Dimensional Quantum Physics, Department of Physics, Tsinghua University, Beijing 100084, China}
\affiliation{Beijing Academy of Quantum Information Sciences, Beijing 100193, China}
\affiliation{Frontier Science Center for Quantum Information, Beijing 100084, China}

\author{Runan Shang}
\affiliation{Beijing Academy of Quantum Information Sciences, Beijing 100193, China}

\author{Xiao Feng}
\affiliation{State Key Laboratory of Low Dimensional Quantum Physics, Department of Physics, Tsinghua University, Beijing 100084, China}
\affiliation{Beijing Academy of Quantum Information Sciences, Beijing 100193, China}
\affiliation{Frontier Science Center for Quantum Information, Beijing 100084, China}

\author{Lin Li}
\affiliation{Beijing Academy of Quantum Information Sciences, Beijing 100193, China}

\author{Qi-Kun Xue}
\affiliation{State Key Laboratory of Low Dimensional Quantum Physics, Department of Physics, Tsinghua University, Beijing 100084, China}
\affiliation{Beijing Academy of Quantum Information Sciences, Beijing 100193, China}
\affiliation{Frontier Science Center for Quantum Information, Beijing 100084, China}
\affiliation{Southern University of Science and Technology, Shenzhen 518055, China}

\author{Ke He}
\email{kehe@tsinghua.edu.cn}
\affiliation{State Key Laboratory of Low Dimensional Quantum Physics, Department of Physics, Tsinghua University, Beijing 100084, China}
\affiliation{Beijing Academy of Quantum Information Sciences, Beijing 100193, China}
\affiliation{Frontier Science Center for Quantum Information, Beijing 100084, China}
\affiliation{Hefei National Laboratory, Hefei 230088, China}

\author{Hao Zhang}
\email{hzquantum@mail.tsinghua.edu.cn}
\affiliation{State Key Laboratory of Low Dimensional Quantum Physics, Department of Physics, Tsinghua University, Beijing 100084, China}
\affiliation{Beijing Academy of Quantum Information Sciences, Beijing 100193, China}
\affiliation{Frontier Science Center for Quantum Information, Beijing 100084, China}


\begin{abstract}

PbTe nanowires coupled to a superconductor have recently been proposed as a potential Majorana platform. The hallmark of the one-dimensional nature of ballistic nanowires is their quantized conductance. Here, we report the observation of conductance plateaus at multiples of the quantized value $2e^2/h$ in PbTe nanowires at finite magnetic fields. The quantized plateaus, as a function of source-drain bias and magnetic field, allow for the extraction of the Land\'e $g$-factor, sub-band spacing and effective mass. The coefficient of 2 in the plateau conductance indicates the presence of valley degeneracy arising from the crystal orientation of the nanowires, which are grown on a (001) substrate. Occasionally, this degeneracy can be lifted by a gate voltage that breaks the mirror symmetry. Our results demonstrate the one-dimensionality of PbTe nanowires and fulfill one of the necessary conditions for the realization of Majorana zero modes.

\end{abstract}

\maketitle  

The conductance of a ballistic one-dimensional electron system is known to be quantized in units of $2e^2/h$ where the coefficient 2 arises from the spin degeneracy. This quantization can be observed in quantum point contacts (QPCs) defined in two dimensional electron gases (2DEGs) \cite{vanWees_1988, Wharam_1988} or semiconductor nanowires \cite{Kammhuber2016, Heedt_2016, Silvano_ballistic, InAs_Gooth, Elham_Cross}. One fascinating application of the one-dimensional electron system is to couple a semiconductor nanowire with a superconductor \cite{Lutchyn2010, Oreg2010}. Majorana zero modes are predicted to emerge and can be further used in topological quantum computations \cite{ReadGreen, Kitaev, RMP_topological}. Much experimental efforts have been devoted to searching for possible Majorana signatures in InAs and InSb nanowires \cite{Mourik, Deng2016, Albrecht, Gul2018, Zhang2021, Song2022, WangZhaoyu,Delft_Kitaev, Prada2020,  NextSteps, Pasquale_Review}. These wires indeed have shown quantized conductance plateaus (at zero or finite magnetic fields) \cite{Kammhuber2016, Heedt_2016}, confirming their one-dimensional nature. After the more-than-a-decade extensive studies on InAs and InSb, PbTe nanowires have recently been proposed as a new Majorana system \cite{CaoZhanPbTe}. The large dielectric constant in PbTe ($\sim$1350) may help to solve the disorder issue which is the current challenge in InAs and InSb \cite{Patrick_Lee_disorder_2012, GoodBadUgly, DasSarma2021Disorder, Tudor2021Disorder}. Soon after, these PbTe nanowires have been grown using selective area epitaxy \cite{Jiangyuying, Erik_PbTe_SAG}, and transport experiments on the Aharonov-Bohm effect \cite{PbTe_AB, Erik_PbTe_SAG} and quantum dots \cite{Fabrizio_PbTe} have been carried out. Furthermore, the superconducting proximity effect has also been demonstrated in PbTe-Pb nanowires, exhibiting a gate-tunable Josephson supercurrent and a hard induced gap \cite{Zitong}. All these improvements are encouraging on establishing PbTe nanowires as a promising Majorana candidate. In this paper, we demonstrate another necessary ingredient, the quantized conductance, confirming the one-dimensionality of PbTe nanowires.  

Figure 1(a) shows a scanning electron micrograph (SEM) of device A. The PbTe nanowire was selectively grown on a CdTe(001) substrate \cite{Jiangyuying}.  Prior to the PbTe growth, the CdTe substrate was covered by a thin Pb$_{1-x}$Eu$_{x}$Te insulating layer, see Fig. S1 in the supplementary material (SM) for details. The value of $x$ is estimated to be $\sim$1$\%$ based on the beam flux. After the wire growth, the PbTe was further capped with a thin layer of Pb$_{1-x}$Eu$_{x}$Te. Source-drain contacts and a side gate were then deposited by evaporating Ti/Au (10/60 nm). The crystal directions and wire orientation are labeled. The measurement was performed in a dilution fridge at its base temperature (below 40 mK) using a standard two-terminal set-up. A series resistance ($R_{\text{series}}$), contributed by the fridge filters ($\sim$3.5 k$\Omega$, not drawn in Fig. 1(b)) and the contact resistance (a few hundred Ohms), were subtracted from the measured two-terminal conductance. The device conductance $G\equiv dI/dV$ was measured as a function of the bias ($V$), gate voltage ($V_{\text{G}}$) and magnetic field ($B$). The bias drop on $R_{\text{series}}$ has also been excluded in $V$. Five devices (A-E) were measured, see Fig. S1 in SM for their SEMs and crystal directions. Devices A, B and C were studied extensively due to their better quantization compared to devices D and E.   

\begin{figure}[htb]
\includegraphics[width=\columnwidth]{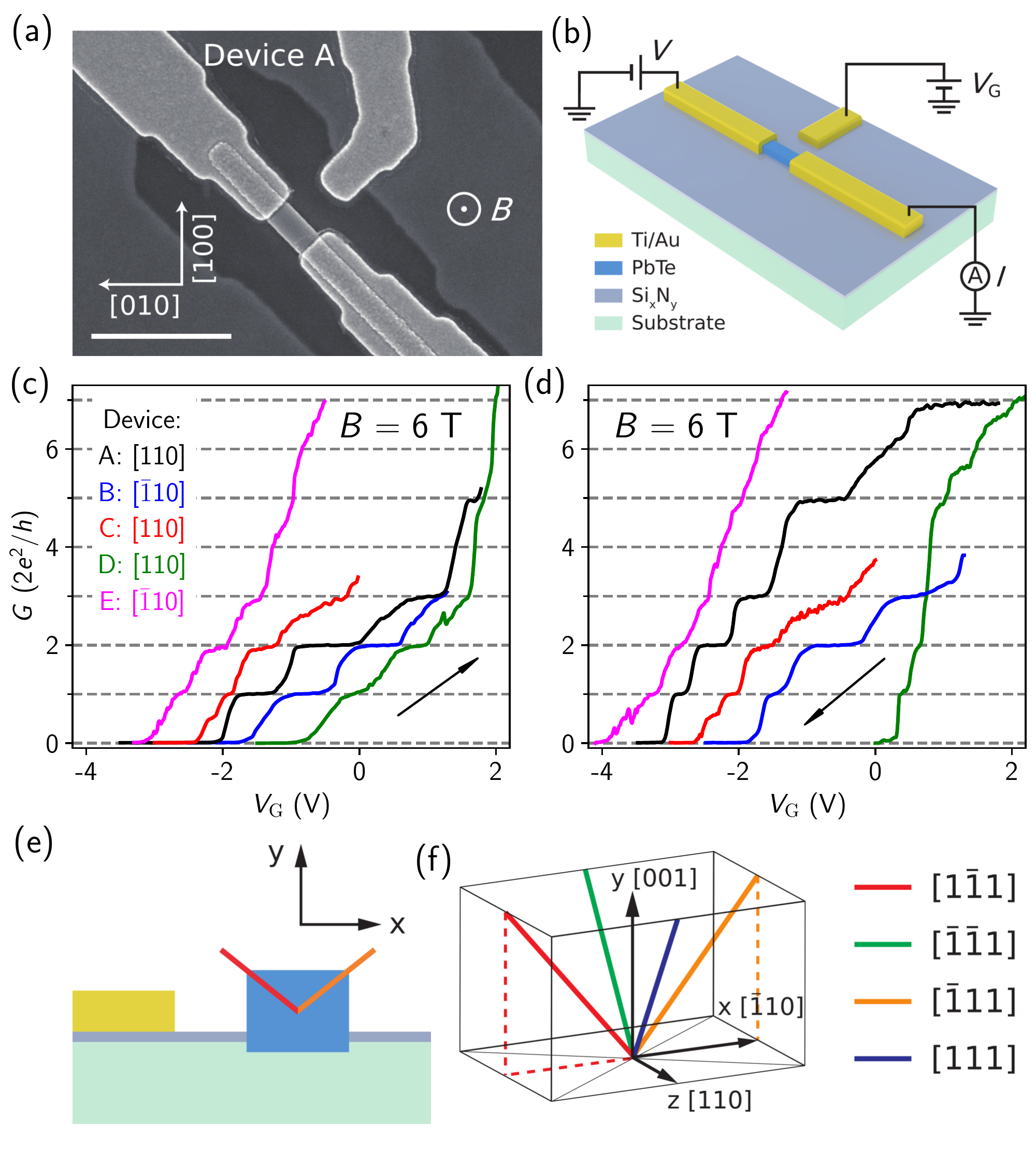}
\centering
\caption{(a) SEM of device A. Scale bar, 1 $\upmu$m. (b) Schematic (simplified) of device A and the measurement circuit. (c) $G$ vs $V_{\text{G}}$ for the five devices. $V$ = 0. $B$ = 6 T. $V_{\text{G}}$ sweeps from negative to positive (see the arrow). (d) Opposite sweeping direction of $V_{\text{G}}$. For clarity, $V_{\text{G}}$ of device D (E) was horizontally shifted by 0.5 (1.5) V in (c) and 2 (0.7) V in (d). (e) Schematic of the device cross-section. (f) 3D illustration of the four valleys. The directions of the [$1\bar{1}1$]-valley (red) and the [$\bar{1}11$]-valley (orange) are also sketched in (e). }
\label{fig1}
\end{figure}

The conductance plateaus at multiples of $2e^2/h$ are clearly visible at a $B$ of 6 T for all five devices, as shown in Figs. 1(c) and 1(d). The $B$-direction was perpendicular to the substrate plane throughout the measurement. The sweeping direction of $V_{\text{G}}$ was from negative to positive for Fig. 1(c) and opposite for Fig. 1(d). Despite the presence of hysteresis, the plateaus are observable for both sweep directions. Previously, quantized conductance in PbTe has been reported in QPCs defined in 2DEGs \cite{PRB_1999_PbTe_QPC, Physica_E_2002_PbTe_QPC, Physica_E_2004_PbTe_QPC, PRB_2005_PbTe_QPC, Physica_E_2006_PbTe_QPC, 2007_Grabecki, Grabecki_2007_JAP}. Compared to nanowires, the plateaus in 2DEG-based QPCs are more robust to disorder-induced scattering, as most of the electrons, after passing through the nano-constriction, are less likely to be scattered back through the constriction again. In nanowires, disorder-induced scattering is more likely to result in back scattering, thereby degrading the plateau quality.

Another notable feature is that four devices (ABDE) show plateaus in units of $2e^2/h$ instead of $e^2/h$. Only device C reveals the $e^2/h$ plateau. Since the spin degeneracy has already been lifted at a $B$ of 6 T (see Fig. 3 for discussion), we interpret the coefficient of  2 coming from the valley degeneracy. This degeneracy is not affected by $B$ since the valley degree of freedom is usually decoupled from $B$. In previous QPC studies on PbTe-based 2DEGs \cite{PRB_1999_PbTe_QPC, Physica_E_2002_PbTe_QPC, Physica_E_2004_PbTe_QPC, PRB_2005_PbTe_QPC, Physica_E_2006_PbTe_QPC, 2007_Grabecki, Grabecki_2007_JAP}, the valley degeneracy is absent (lifted)  due to the (111) growth direction, consistent with the theory calculation \cite{CaoZhanPbTe}. In our case, the substrate is along the (001) direction and the nanowire axis is along the [110] direction (for devices ACD) or [$\bar{1}10$] direction (for devices BE). Theory suggests that the valley-degeneracy cannot be fully lifted for wires along these directions \cite{CaoZhanPbTe}. The orientations of the four valleys are sketched in Figs. 1(e) and 1(f) for device A. The [$\bar{1}$$\bar{1}1$]-valley (green) and the [111]-valley (blue) are degenerate while the [$1\bar{1}1$]-valley (red) and the [$\bar{1}11$]-valley (orange) are degenerate \cite{CaoZhanPbTe}. This double two-fold degeneracy could explain the observed plateaus in steps of $2e^2/h$.

The degeneracy between the ``red'' and ``orange'' valleys can be lifted by breaking the mirror symmetry about the y-axis, e.g. by covering a side facet of the nanowire with a superconductor \cite{CaoZhanPbTe}. The work-function mismatch between the superconductor and the nanowire leads to charge transfer and the build-up of an electric field. This field breaks the y-axis mirror symmetry.  Alternatively, a side gate in principle could also induce an electric field and lift this degeneracy (e.g. the case of device C). This electric field, however,  is typically weak due to the screening effect of the large dielectric constant of PbTe. Therefore, the side gate may primarily tune the electro-chemical potential (carrier density) rather than break the symmetry, likely the cases for devices A, B, D, and E. Even if the degeneracy between the ``red'' and ``orange'' valleys is lifted, the degeneracy between the ``green'' and ``blue'' valleys still remains. And if the first few sub-bands are associated with the ``green'' and ``blue'' valleys, the corresponding plateaus would still exhibit steps of $2e^2/h$. In addition to this double two-fold valley degeneracy, the missing steps at $4\times 2e^2/h$ and $6\times 2e^2/h$ for device A suggests the presence of additional valley or orbital degeneracy.  

\begin{figure}[b]
\includegraphics[width=\columnwidth]{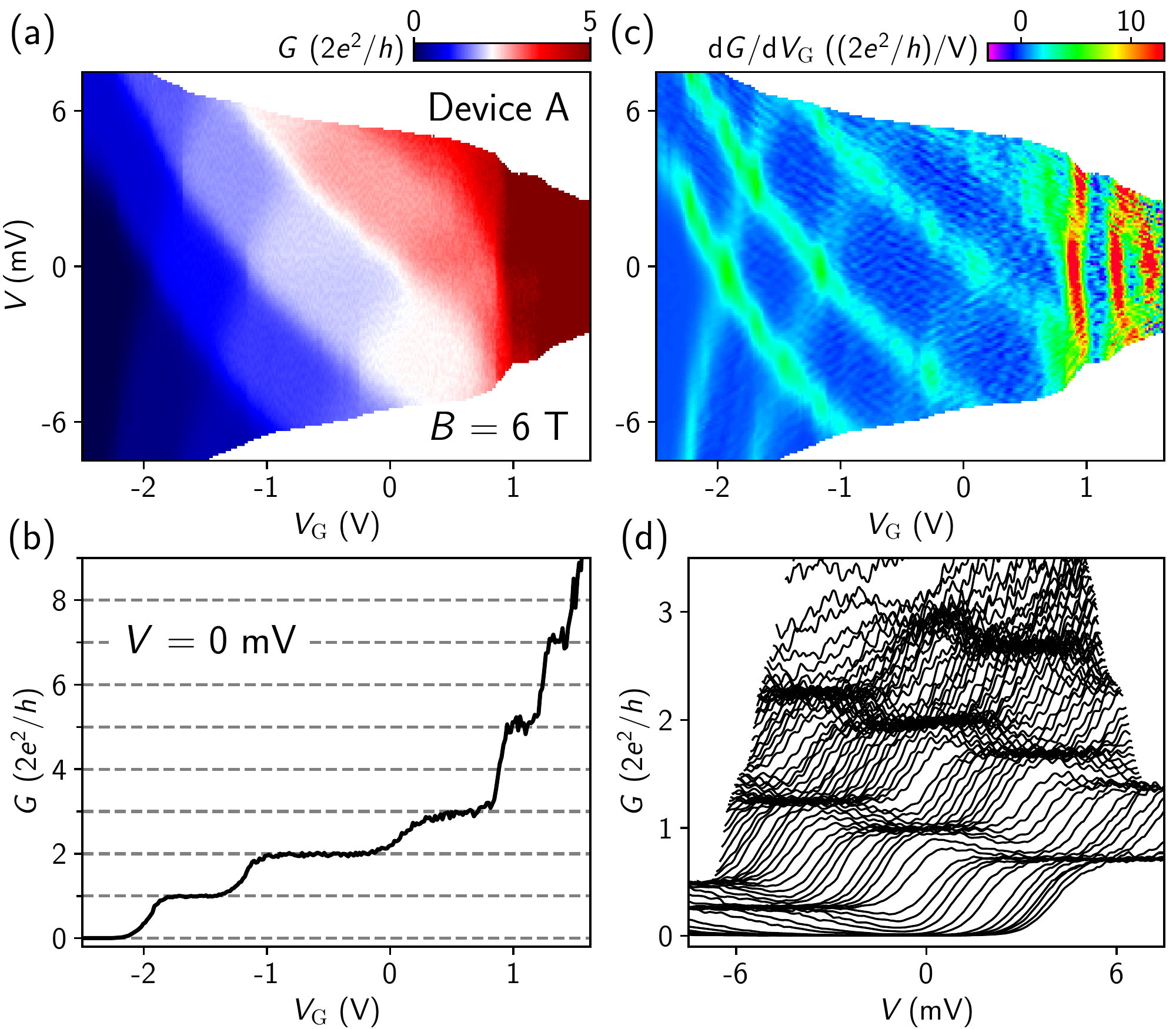}
\centering
\caption{(a) $G$ vs $V$ and $V_{\text{G}}$ at $B$ = 6 T for device A. (b) Zero-bias line cut. (c) Trans-conductance ($dG/dV_{\text{G}}$) of (a). A minor smoothing (``low-pass'', averaging window of 2)  along $V_{\text{G}}$ was performed for clarity. (d) Waterfall plot of (a). Each curve corresponds to a fixed $V_{\text{G}}$.  }
\label{fig2}
\end{figure}

Figure 2(a) shows the 2D color map of the conductance ($G$) in the ($V$, $V_{\text{G}}$) space of device A. The $V_{\text{G}}$ sweep direction was from negative to positive for all the color maps (in Figs. 2-4). The plateaus are resolved as diamond-shaped regions. The diamond sizes (in $V$) match the energy spacing between different sub-bands. The zero-bias line cut (Fig. 2(b)) reveals plateaus of 1, 2, 3, 5 and 7 (in units of $2e^2/h$). The 4, 6 and 8 plateaus are either weak or barely visible, indicating additional degeneracies. The high-bias diamonds correspond to the ``fractional plateaus'' (see Fig. S2 in SM for line cuts). This is not due to the lifting of degeneracy but purely a biasing effect: the quasi-Fermi levels of the source and drain contacts are aligned with different spin-resolved sub-bands.

The plateau diamonds can be better resolved in the ``transconductance'' (Fig. 2(c)), by performing the derivative $dG/dV_{\text{G}}$. The plateau regions correspond to the nearly zero transconductance while the transitions between plateaus show high transconductance. The diamond shapes are tilted, possibly due to the ``cross-talk'' between $V$ and $V_{\text{G}}$. The bias $V$ may act as a ``leakage gate'', thereby $V_{\text{G}}$ needs adjustment to compensate for the effect of $V$, leading to the tilted shapes.

Figure 2(d) presents the line cuts of Fig. 2(a) using the ``waterfall'' plot. Plateaus now are resolved as clusters of line cuts (the dense region). The fractional plateau values at high bias are not halves of those of the zero-bias plateaus, possibly due to the asymmetric bias of the measurement circuit: The lock-in excitation $dV$ is not symmetrically shared by the quasi-Fermi levels of the source and drain contacts, i.e. not $(dV)/2$ for each.

The energy of the lowest sub-band ($E_{1\uparrow}$ and $E_{1\downarrow}$) can be affected by $B$ through its orbital and Zeeman effect. The up and down arrows indicate the spin directions. Note that $E_{1\uparrow}$ or $E_{1\downarrow}$ still has the valley-degeneracy. At finite $B$'s, the orbital effect for $E_{1\downarrow}$ and $E_{1\uparrow}$ is the same, therefore the sub-band spacing ($E_{1\downarrow}-E_{1\uparrow}$) reflects the Zeeman splitting, $g\mu_{\text{B}}B$ (see Fig. 3(d) for the sub-band schematic). $g$ is the Land\'e $g$-factor and $\mu_{\text{B}}$ is the Bohr magneton. Based on the diamond size of the first plateau, $\sim$3.6 meV, we estimate the effective $g$-factor, $g \sim$10. The diamond sizes of the second and third plateaus correspond to the sub-band spacing of $E_{2\uparrow}-E_{1\downarrow} \sim4.16$ meV and $E_{2\downarrow}-E_{2\uparrow} \sim2.83$ meV, respectively. The $g$-factor of the $E_2$ sub-band is estimated to be $\sim$8. These values are significantly smaller than that of the numerical study  \cite{CaoZhanPbTe}, where the $g$-factor is calculated to be $\sim$36 for all the four valleys for this $B$-direction (y-axis in Fig. 1(e)). This $g$-factor reduction is probably related to the confinement \cite{g-factor_reduction}.

To further study the Zeeman effect, in Fig. 3 we show the $B$-evolution of the plateaus in device A. At zero field, no plateaus can be resolved due to residue disorder in the device. At higher fields, the electron backscattering is significantly suppressed by the Lorentz force and plateaus start to emerge (see Fig. 3(b) for line cuts). The first plateau is observable for $B >$ 4 T, and its size scales roughly linearly with $B$, consistent with the interpretation of Zeeman splitting.

\begin{figure}[tb]
\includegraphics[width=\columnwidth]{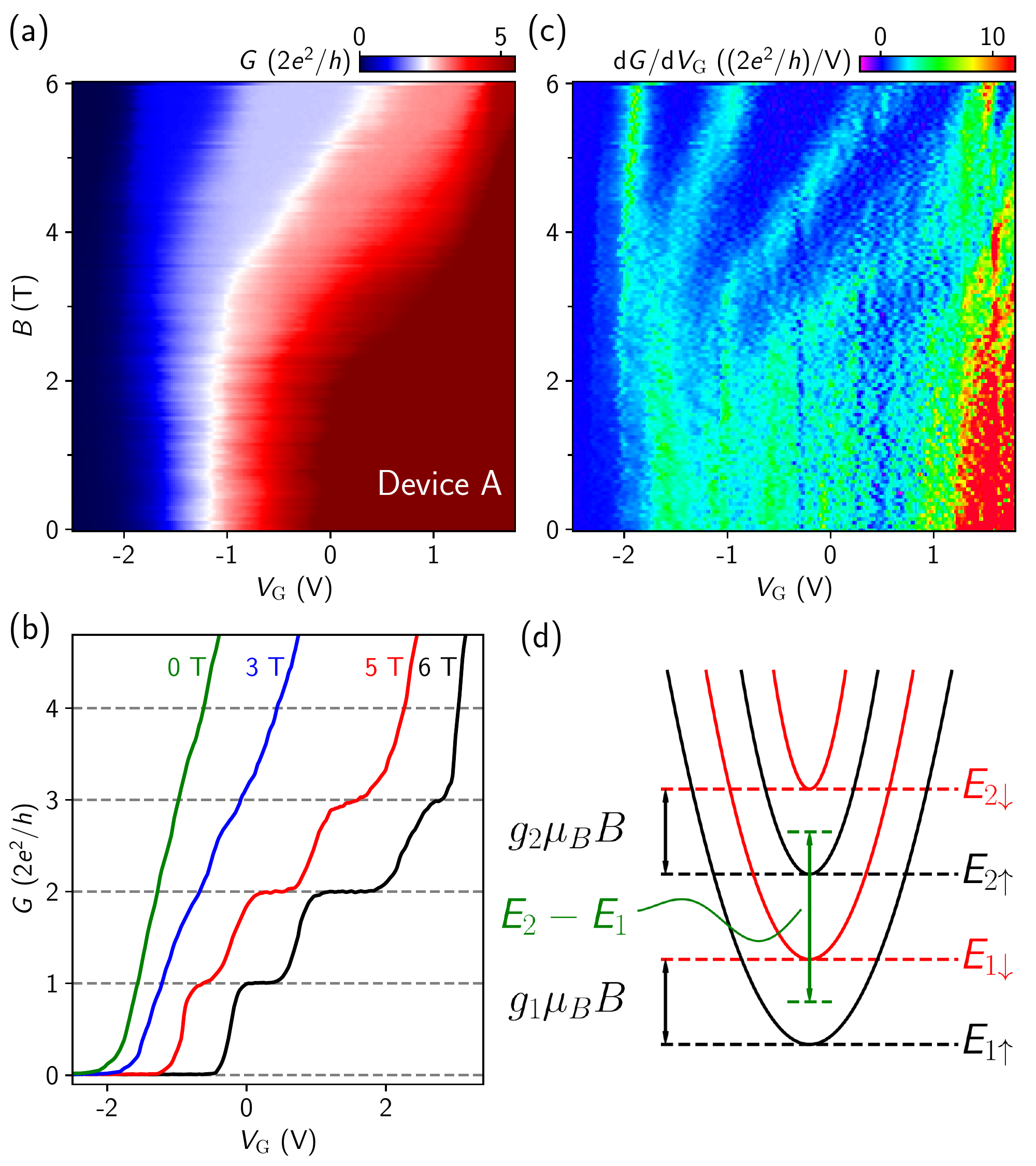}
\centering
\caption{(a) Zero-bias $G$ vs $B$ and $V_{\text{G}}$ for device A. (b) Several line cuts of (a). The curves are offset horizontally (by 0.5 V between each curve) for clarity. (c) Transconductance of (a). (d) Sketch of the lowest four spin-resolved sub-bands. }
\label{fig3}
\end{figure}

The transitions between plateaus (peaks in Fig. 3(c)) have ``bended'' curvatures (instead of linear) in $B$, indicating the presence of orbital effect. The sub-band spacing (without the Zeeman effect) is $E_2-E_1 = \hbar \omega(B)=\hbar\sqrt{\omega_0^2+\omega_c^2}$. $\omega_0$ is associated with the confinement potential $\frac{1}{2}m^*\omega_0^2x^2$ at zero field. Note that we neglect the electron motion along the y-axis (see Fig. 1(e)) since the wire thickness ($\sim$90 nm) is significantly smaller than its width ($\sim$200 nm). $\omega_c=eB/m^*$ is the cyclotron frequency. The sub-band spacing at finite $B$ (without Zeeman) is a combined effect of the nanowire confinement ($\omega_0$) and the magnetic confinement ($\omega_c$, the orbital effect), also known as the magnetic depopulation \cite{1991_PRB_QPC}. Now taking into account the Zeeman effect, we define $E_1=(E_{1\uparrow}+E_{1\downarrow})/2$, and $E_2=(E_{2\uparrow}+E_{2\downarrow})/2$. Then $\hbar\omega(B)=(E_{2\downarrow}-E_{2\uparrow})/2+E_{2\uparrow}-E_{1\downarrow}+(E_{1\downarrow}-E_{1\uparrow})/2=1/2(g_1+g_2)\mu_{\text{B}}B+E_{2\uparrow}-E_{1\downarrow}$. $g_1$ (= 10) and $g_2$ (= 8) are the $g$-factors of the $E_1$ and $E_2$ sub-bands, see Fig. 3(d) for the illustration.  We then fit the size of the second plateau ($E_{2\uparrow}-E_{1\downarrow}$) with the formula above based on Fig. 3(c) ($V_{\text{G}}$ is converted to energy based on Fig. 2). An effective mass, $m^* \sim0.09m_{\text{e}}$, can be estimated. $m_{\text{e}}$ is the free electron mass. This value (0.09) of the effective mass is between the value of the green/blue-valley (0.168) and the red/orange-valley (0.024) from the numerical calculation \cite{CaoZhanPbTe}. It should be noted that the estimation is rough and based on a simplified model. For example, it assumes that the $E_1$ and $E_2$ sub-bands have the same effective mass which may not be the case as reflected by the $g$-factor difference. Moreover, the anisotropy of the effective mass and band diagram can bring additional modifications.

\begin{figure}[tb]
\includegraphics[width=\columnwidth]{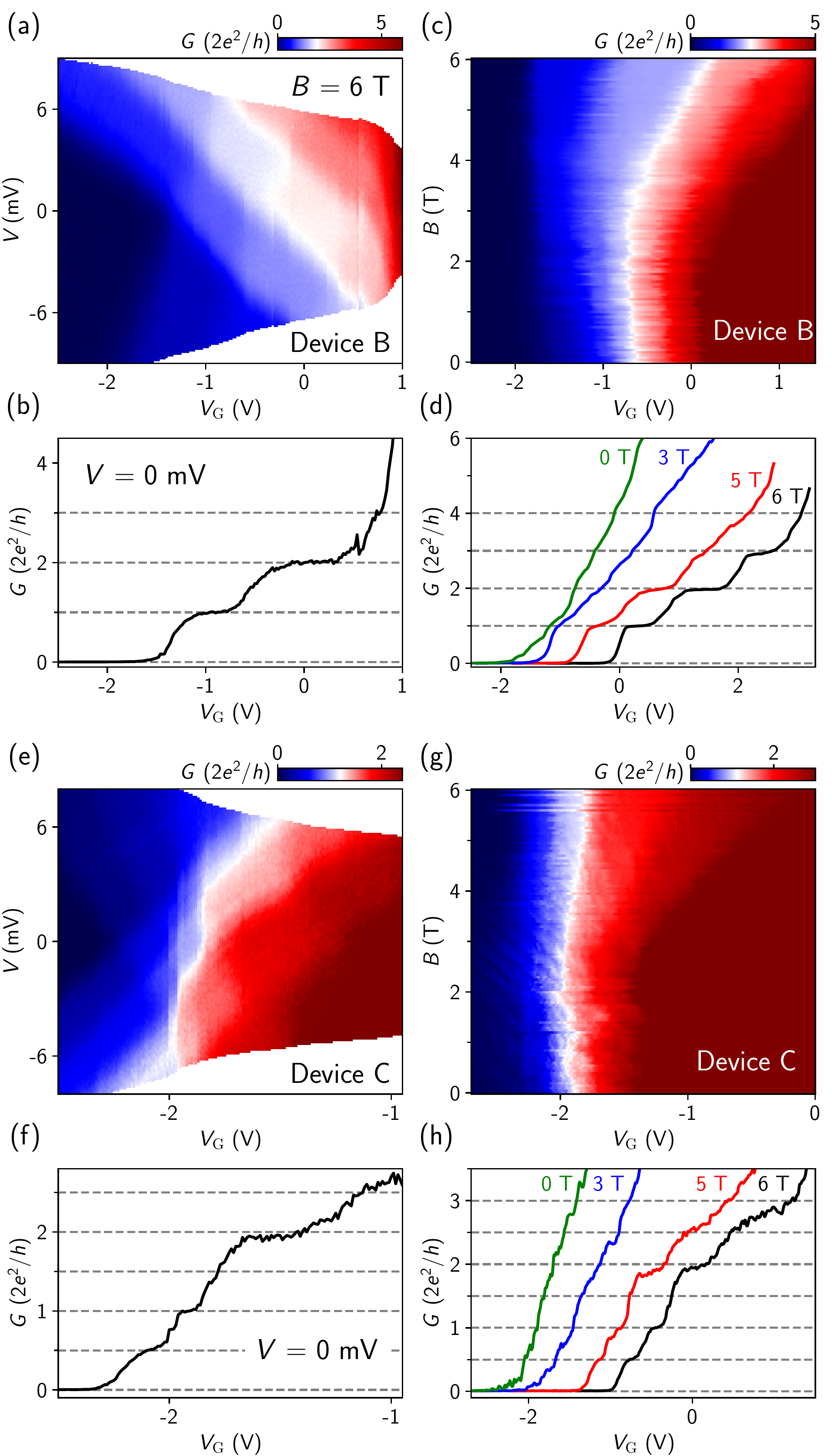}
\centering
\caption{(a) $G$ vs $V$ and $V_{\text{G}}$ for device B at $B$ = 6 T. (b) Zero-bias line cut. A charge jump is visible between the second and the third plateau due to the device instability. (c) Zero-bias $G$ vs $B$ and $V_{\text{G}}$ for device B. (d) Line cuts from (c) with horizontal offsets (by 0.6 V between each curve) for clarity. (e) $G$ vs $V$ and $V_{\text{G}}$ for device C at $B$ = 6 T. (f) Zero-bias line cut. (g) Zero-bias $G$ vs $B$ and $V_{\text{G}}$ for device C. (h) Line cuts from (g) with horizontal offsets (0.5 V for each curve) for clarity. }
\label{fig4}
\end{figure}

In Fig. 4, we show the bias and $B$ scans of two more devices (B and C). For device B, the diamonds for the first two plateaus are clearly visible (Fig. 4(a)). Based on the first plateau size ($\sim$3.53 meV), a $g$-factor of $\sim$10 is estimated. The $B$-scan in Fig. 4(d) can resolve three plateaus, see Fig. S3 in SM for the transconductance. For device C, a small $e^2/h$ plateau can be resolved in Figs. 4(e) and 4(f). A possible explanation is the lifting of the degeneracy between the [$1\bar{1}1$]-valley and the [$\bar{1}11$]-valley. This lifting could be caused by the side gate or impurities in device C. The quantization quality of device C is worse than that in devices A and B, suggesting a higher level of disorder. The transconductance (Fig. S3(d)) of the plateau evolution in $B$ suggests that the $e^2/h$ and $2e^2/h$ plateaus together reflect the Zeeman splitting, and the $e^2/h$-plateau is likely the lifting of valley-degeneracy. Based on this interpretation, a $g$-factor of $\sim$9 can be estimated. The small kink near $0.75\times2e^2/h$ in Fig. 4(f) is likely a charge jump which is absent in Fig. 4(h) (the black curve). 

The valley degeneracy in PbTe nanowires is ``unwanted'' for Majorana research. Both theory \cite{CaoZhanPbTe} and experiments \cite{PRB_1999_PbTe_QPC, Physica_E_2002_PbTe_QPC, Physica_E_2004_PbTe_QPC, PRB_2005_PbTe_QPC, Physica_E_2006_PbTe_QPC, 2007_Grabecki, Grabecki_2007_JAP, Fabrizio_PbTe} have demonstrated that the valley degeneracy can be lifted for PbTe nanowires grown on a (111) substrate. Future works on hybrid superconductor-semiconductor nanowires could focus on this crystal direction. The (001) substrate (the case in this paper) could still be an option for Majorana research, as a superconductor grown on a side facet can induce an electric field and lift the valley degeneracy \cite{CaoZhanPbTe}. Moreover, the ubiquitous valley-degeneracy of PbTe wires on (001) substrate may be useful for studying other intriguing physics. For example, it may provide a material platform to observe the SU(4) (or SU(8)) Kondo effect \cite{SU4_PRL_2002, SU4_PRL_2007, SU4_NP_2014} in PbTe quantum dots.

In summary, we have observed quantized conductance plateaus in PbTe nanowires at non-zero magnetic fields. The land\'e $g$-factor, sub-band spacing, effective mass, valley degeneracy and the lifting of the degeneracy are further discussed. Our results demonstrate the one-dimensional nature of PbTe nanowires and fulfill one necessary requirement for future Majorana researches based on this material. Besides the hybrid devices with a superconductor, future work on PbTe nanowires could aim to (1) further reduce the disorder level to achieve conductance quantization at zero magnetic field; (2) search for cleaner signatures of the spin-orbit (helical) gap \cite{PRB_2014_helical_gap}; (3) explore other nanowire crystal orientations to investigate different valley-degeneracy, degeneracy-lifting and $g$-factors. Apart from Majorana-related research, the excellent gate-tunability of these PbTe nanowires could also be utilized for the realization of spin qubits \cite{Petta_Science}, superconducting qubits (gatemons) \cite{2015_PRL_gatemon,DiCarlo_gatemon, Huo_gatemon} or other quantum devices like Cooper-pair-splitters \cite{Nature_Cooper_pair}. 

\textbf{Acknowledgment} This work is supported by Tsinghua University Initiative Scientific Research Program, National Natural Science Foundation of China (92065206) and the Innovation Program for Quantum Science and Technology (2021ZD0302400). Raw data and processing codes within this paper are available at https://doi.org/10.5281/zenodo.7848599.

\bibliography{mybibfile}

\newpage

\onecolumngrid

\newpage


\section*{Supplementary material}

# Supplementary Materials for "Conductance Quantization in PbTe Nanowires"

Wenyu Song,* Yuhao Wang,* Wentao Miao, Zehao Yu, Yichun Gao, Ruidong Li, Shuai Yang, Fangting Chen, Zuhan Geng, Zitong Zhang, Shan Zhang, Yunyi Zang, Zhan Cao, Dong E. Liu, Runan Shang, Xiao Feng, Lin Li, Qi-Kun Xue, Ke He,† and Hao Zhang‡

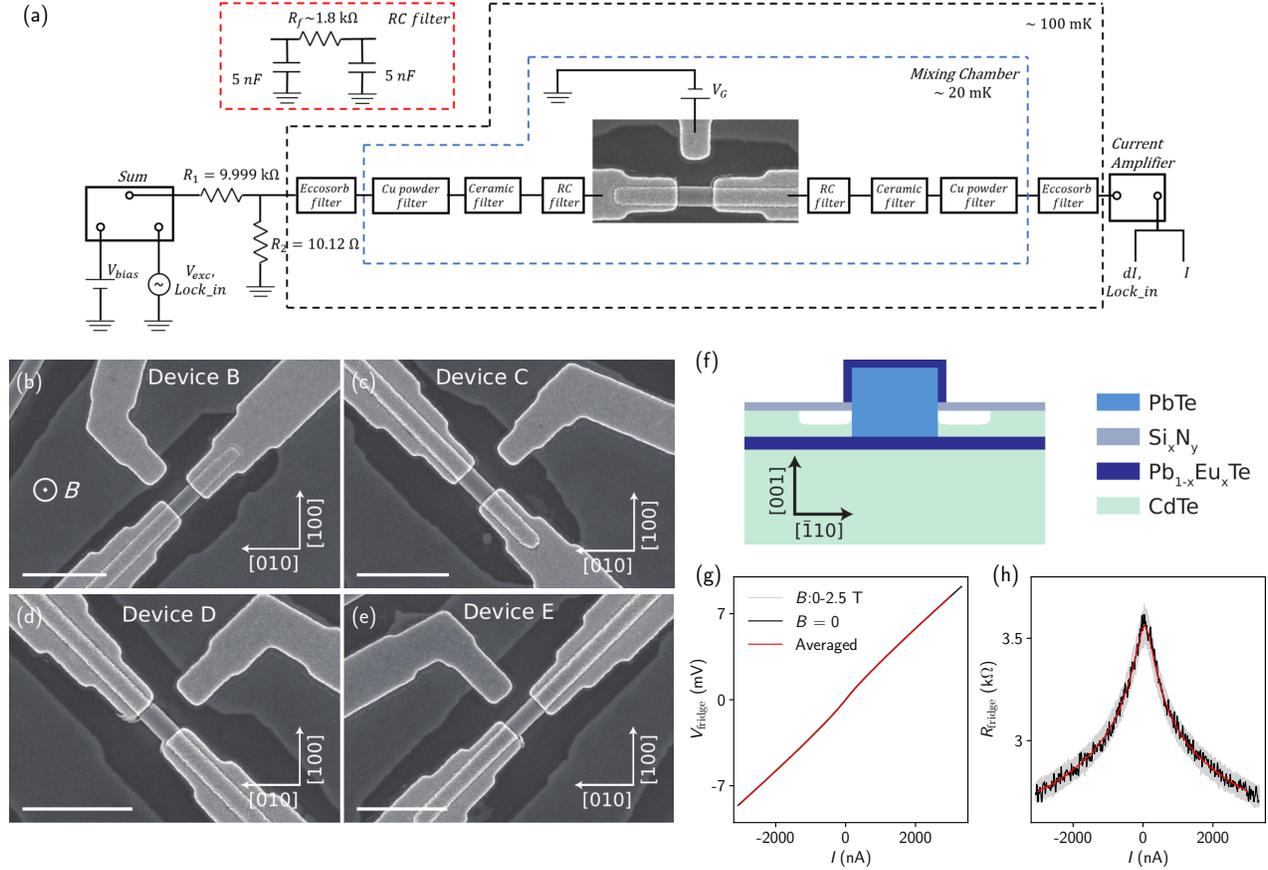

FIG. S1. (a) Measurement circuit diagram. A DC voltage $V_{\text{bias}}$ together with a lock-in excitation $V_{\text{ac}}$ passes through a voltage divider, the three-stage filters and then connects to the source contact of the device. The current $I$ and $dI$ were measured from the drain. (b-e): SEMs of devices B, C, D and E. Scale bar, 1 μm. (f) Schematic of the PbTe nanowire cross-section. (g) Calibration of the fridge $I$-$V$ curve by shorting two fridge lines at the mixing chamber. (h) The fridge differential resistance measured using the same method. The signals are mainly contributed by the filters. The black curve was measured at $B = 0$ T. The gray curves were measured for $B$'s between 0 and 2.5 T. The red curve is the average. The data of (g) and (h) have been used in our previous work (arXiv: 2212.02132) to calibrate devices there. $R_{\text{series}} = R_{\text{fridge}} + R_{\text{contact}}$. $R_{\text{contact}}$ is a fitting parameter for each device to fit the plateaus to the quantized values. $R_{\text{contact}}$ is 520, 800, 600 and 900 Ohms for devices B, C, D and E, respectively. For device A, $R_{\text{contact}}$ is 1050 Ohms for Figs. 1(c), 1(d) and 3. For Figs. 2 and S2(a), it is 900 Ohms. This 150-Ohm difference is close to the error bar of $R_{\text{fridge}}$ (see (h)) and may originate from the slightly different base temperatures or bias offsets between those measurements. The bias drop on the device, $V = \alpha V_{\text{bias}} - V_{\text{fridge}}(I) - R_{\text{contact}} \times I - V_{\text{offset}}$. $\alpha = 1/1011$ is the voltage ratio of the divider (including the output resistance of the summing module). $V_{\text{fridge}}(I)$ is the voltage drop on the fridge at current $I$ (see panel (g)). The device conductance $G \equiv dI/dV = 1/((\alpha V_{\text{ac}}/dI - R_{\text{fridge}}(I) - R_{\text{contact}}))$. $R_{\text{fridge}}(I)$ is interpolated from (g) (lower panel). In this way, $G(V)$ has excluded the contributions of $R_{\text{series}}$ in both $G$ and $V$.

* equal contribution
† kehe@tsinghua.edu.cn
‡ hzquantum@mail.tsinghua.edu.cn

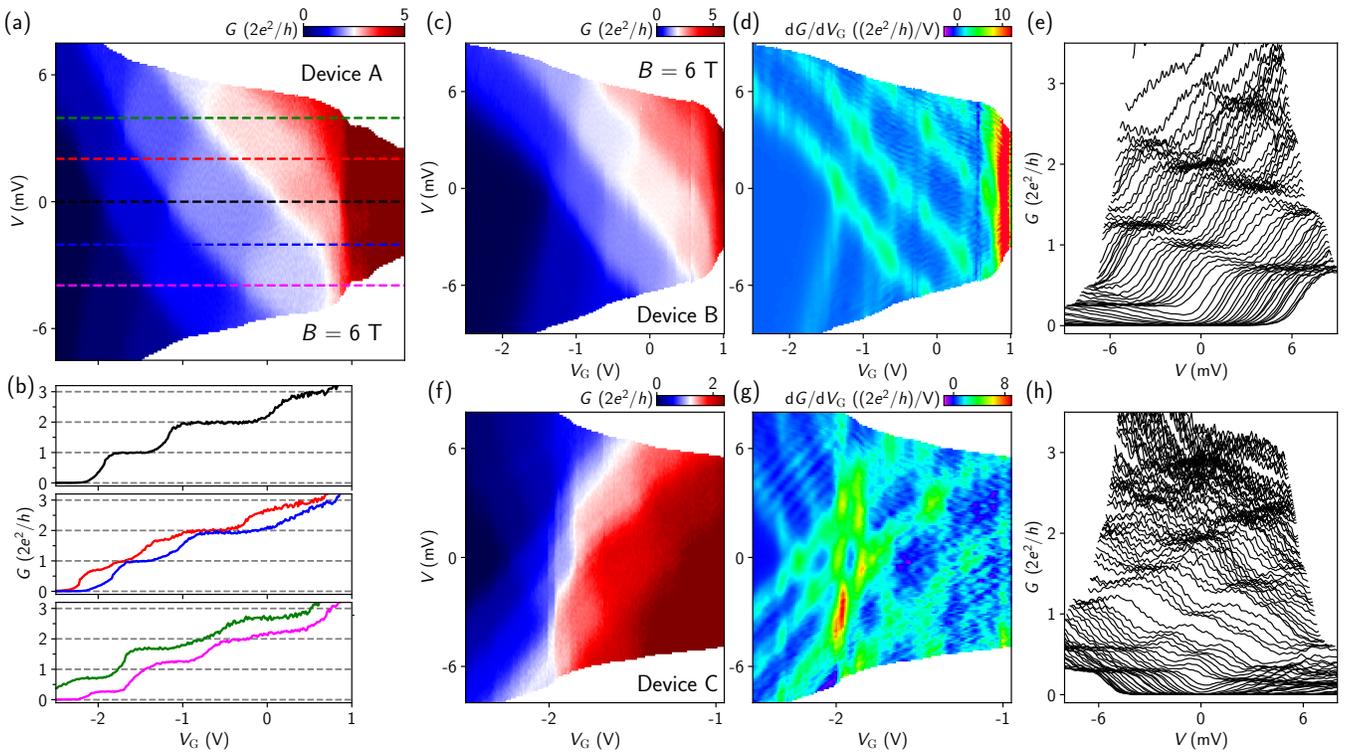

FIG. S2. (a) Re-plot of Fig. 2(a). (b) Zero-bias line cut (the black curve) and finite bias line cuts (see the dashed lines of corresponding colors in (a)). The values of the fractional plateaus at high bias are not at the halves of those of the zero-bias, possibly due to the asymmetric bias. (c) Re-plot of Fig. 4(a). (d) Transconductance of (c). A minor smoothing (low-pass, averaging window of 2) is performed for $V_G$. (e) Waterfall plots of (c). (f)-(h) Similar to (c)-(e) but for device C.

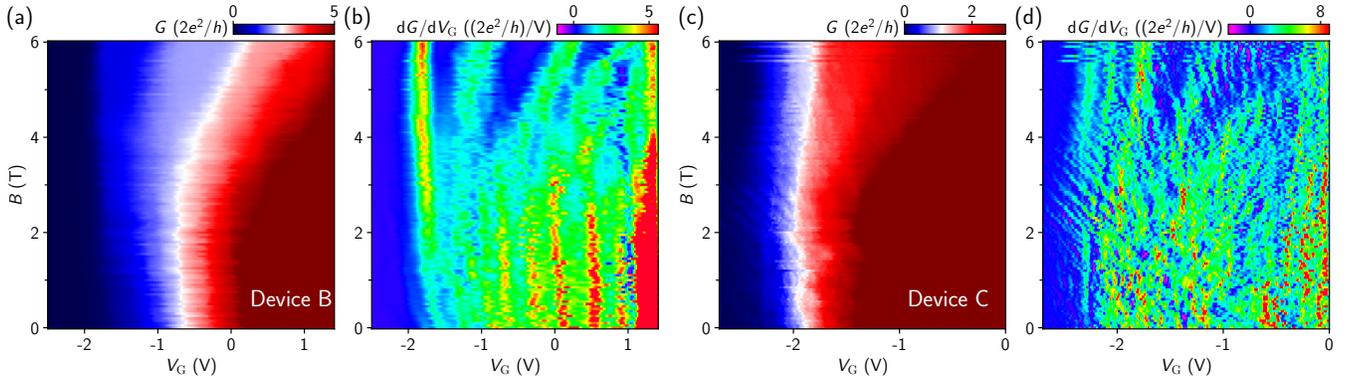

FIG. S3. (a) Re-plot of Fig. 4(c). (b) Transconductance of (a). (c) Re-plot of Fig. 4(g). (d) Tranconductance of (c). A minor smoothing (low-pass, averaging window of 2) is performed for $V_G$ for (b) and (d).



\end{document}